\documentclass[12pt,a4paper]{article}

\usepackage{algorithm}
\usepackage{algpseudocode}
\usepackage{geometry}
\usepackage[utf8]{inputenc}
\usepackage[T1]{fontenc}
\RequirePackage{amsmath,amssymb}
\RequirePackage{mathrsfs}
\RequirePackage{amsfonts,bm}
\RequirePackage{dsfont}
\RequirePackage{braket}
\RequirePackage{tabularx}
\usepackage{multirow}
\RequirePackage{nicefrac}
\RequirePackage{graphicx}
\RequirePackage{booktabs}
\RequirePackage{colortbl}
\RequirePackage{xcolor}
\RequirePackage[symbol]{footmisc}
\usepackage{mathtools}
\usepackage{comment}
\usepackage{blkarray}
\usepackage{float}
\usepackage{rotating}
\usepackage{hhline}
\usepackage{tikz}
\usepackage{multirow}
\usepackage[section]{placeins}
\usepackage[natbib=true,maxbibnames=99,giveninits=true,sorting=none]{biblatex}
\newenvironment{aline}
    {\begin{equation}
    \begin{aligned}
    }
    { 
    \end{aligned}
    \end{equation}
    \ignorespacesafterend
    }

\def\beq{\begin{equation}}
\def\eeq{\end{equation}}
\def\beqn{\begin{eqnarray}}
\def\eeqn{\end{eqnarray}}

\usepackage{empheq}
\usepackage[most]{tcolorbox}
\usepackage{blkarray}
\newtcbox{\mymath}[1][]{%
    nobeforeafter, math upper, tcbox raise base,
    enhanced, colframe=blue!30!black,
    colback=blue!30, boxrule=1pt,
    #1}

\newcommand{\CC}[2]{C{#1\atopwithdelims[]#2}}

\newcommand{\ba}{\begin{eqnarray}}
\newcommand{\ea}{\end{eqnarray}}

\newcommand{\vth}{\vartheta} 
\newcommand{\vthb}{\bar{\vartheta}} 
\newcommand{\smb}[2]{\hspace{-0.02cm}\left[\begin{smallmatrix}#1\\#2\end{smallmatrix} \right]}
\newcommand{\smbar}[4]{\hspace{-0.02cm}\left[\begin{smallmatrix}#1\\#2\end{smallmatrix} \middle| \begin{smallmatrix}#3\\#4\end{smallmatrix}  \right]}

\newcommand{\one}{\mathds{1}} 
\newcommand{\Sv}{\bm{S}} 
\newcommand\e[1]{\bm{e_#1}}
\newcommand\bv[1]{\bm{b_#1}}
\newcommand\z[1]{\bm{z_#1}}

\newcommand\tr{\text{tr}}

\numberwithin{equation}{section}

\begin{document}
\begin{titlepage}
\samepage{
\setcounter{page}{1}
\rightline{LTH-1332}
\rightline{February 2023}

\vfill
\begin{center}
  {\Large \bf{
     Fayet--Iliopoulos D--Term \\ \medskip 
     in Non--Supersymmetric Heterotic String Orbifolds }}

\vspace{1cm}
\vfill

{\large Alonzo R. Diaz Avalos $^{1}$\footnote{E-mail address: a.diaz-avalos@liverpool.ac.uk}, \\ \medskip
Alon E. Faraggi$^{1,2}$\footnote{E-mail address: alon.faraggi@liverpool.ac.uk}, Viktor G. Matyas$^{1}$\footnote{E-mail address: viktor.matyas@liverpool.ac.uk} and 
 Benjamin Percival$^{1}$\footnote{E-mail address: b.percival@liverpool.ac.uk}}

\vspace{1cm}

{\it $^{1}$ Dept.\ of Mathematical Sciences, University of Liverpool, Liverpool
L69 7ZL, UK\\}
\vspace{.08in}

{\it $^{2}$ CERN, Theoretical Physics Department, CH--1211 Geneva 23, Switzerland\\}

\vspace{.025in}
\end{center}

\vfill
\begin{abstract}
\noindent
The Fayet--Iliopoulos $D$--term is a common feature in $\mathcal{N}=1$ string vacua that contain an anomalous $U(1)$ gauge symmetry, and arises from a one--loop diagram in string perturbation theory. The 
same diagram is generated in string vacua in which supersymmetry is broken directly at the 
string scale, either via spontaneous Scherk--Schwarz breaking, in which case the gravitino 
mass is determined by the radius of the circle used in the Scherk--Schwarz mechanism, or via explicit supersymmetry breaking by the GSO projections. We analyse the resulting would--be 
Fayet--Illiopoulos $D$--term in the non--supersymmetric string vacua and its contribution to the
vacuum energy. A numerical estimate in an explicit tachyon--free string--derived model suggests that 
the would--be $D$--term contribution may uplift the vacuum energy to a positive value.

\end{abstract}

\smallskip}

\end{titlepage}

\section{Introduction}\label{intro}

String theory provides the most advanced framework to explore how 
the Standard Model parameters may arise from a fundamental theory
of the gauge and gravitational interactions. 
Toward this end, string models that reproduce the spectrum of the
Minimal Supersymmetric Standard Model were constructed 
\cite{fny, slm2, cfn, slmclass, Lebedev:2006kn, Blaszczyk:2010db}.
Amongst them, the free fermionic Standard--Like Models (SLMs)
\cite{fny, slm2, cfn, slmclass} 
are some of the most studied examples.
The heterotic string in particular provides a compelling
framework to study the gauge--gravity unification as it
reproduces the embedding of the Standard Model chiral spectrum
in spinorial $SO(10)$ multiplets.
Three generation heterotic string models with $SO(10)$ embedding of the
Standard Model charges have been constructed since the late eighties.
While the early constructions consisted of isolated examples,
over the past two decades, systematic computerised methods have been
developed that enable the exploration of large spaces of vacua and the
extraction of their main characteristics. 
This methodology led to the important discoveries of spinor--vector
duality \cite{fkr2} and exophobic vacua \cite{acfkr1}. 

The majority of the phenomenological string models constructed to date
possess $\mathcal{N}=1$ spacetime supersymmetry, 
whereas non--supersymmetric vacua have been studied sporadically
\cite{AGMV, so10tclass, PStclass, Saul, asymmclass, DH, ADM, aafs, CoCSuppression2, nonsusy5, stable, 
NSUSYBranes4}. 
The advantage of supersymmetric backgrounds
is that they are stable and some of the properties
inferred from them, {\it e.g.} the number of chiral generations and their
charges, are certain. It is clear, however, that addressing
many of the open questions in string phenomenology mandates the exploration
of non--supersymmetric configurations. In particular, those pertaining to the
dynamical and cosmological evolution close to the Planck scale. The key
issue there is the instability of the non--supersymmetric vacua, 
which generically give rise to tachyonic instabilities. However, even those
that are free of physical tachyons, have non--vanishing vacuum energy
and other non--vanishing tadpole amplitudes and are therefore, in general,
unstable.

One of the prevailing features of the supersymmetric string models
is the existence of an anomalous $U(1)$, which is cancelled by an analogue
of the Green--Schwarz mechanism. The anomalous $U(1)$ generates a
Fayet--Iliopoulus $D$--term that breaks supersymmetry, which can be
restored by assigning vacuum expectation values to some fields in the string
spectrum along $F$-- and $D$--flat directions.

The Fayet--Iliopoulos $D$--term is generated by a one--loop tadpole diagram
in string perturbation theory whenever there exists an anomalous $U(1)$ in the
spectrum. In the presence of an anomalous $U(1)$ 
the diagram is present also in string vacua in which supersymmetry
is broken directly at the string scale and contributes to the non--vanishing
vacuum energy. This contribution is particularly relevant for the question of
whether de Sitter vacua exist in string theory since a positive would--be $D$--term
contribution may dominate a negative vacuum energy contribution
and produce a vacuum with a positive one--loop cosmological constant \cite{BKQ}. The would--be
$D$--term may also play a role in scenarios of $D$--term inflation \cite{Halyo}.

In this paper we, therefore, undertake the task of calculating the
would--be $D$--term in non--supersymmetric string vacua. In Section
\ref{AnomU1} the calculation is set up and carried out by using the
background field method 
\cite{Marios}. 
We then perform a scan of chiral non--supersymmetric string
vacua with unbroken $SO(10)$ symmetry using the systematic
free fermion classification method. The computerised analysis
ensures that the vacua are free of physical tachyons at the free fermionic point and calculates the traces of the $U(1)$ symmetries, hence extracting
the vacua with an anomalous $U(1)$. We then perform a comparative
investigation of the one--loop vacuum amplitude with respect to the
anomalous $U(1)$ would--be $D$--term contributions.
Following Florakis and Rizos \cite{fr1}, we perform a numerical analysis of the potential for a specific
string model as a function of the moduli in the vicinity of a local minimum  
and suggest that the would--be $D$--term contribution may indeed uplift the 
vacuum energy to a positive value. In this numerical analysis, the string model is found to be tachyon free for any value of the 
moduli in which the potential is being varied by. In this paper, we illustrate the $D$--term uplifting mechanism in 
a single exemplary model. A more extensive discussion, with further examples, will
be given in a forthcoming publication \cite{DFMP2}. 


\section{Anomalous $U(1)$ Tadpole Calculation}\label{AnomU1}

In a four--dimensional heterotic theory the gauge group may contain some $U(1)$ symmetries which are anomalous, namely the sum of the $U(1)$ charges is not zero. In four dimensions the anomalies come from the triangle diagrams, with $U(1)$ fields or $U(1)$ mixed with gravitons as external legs. Under a $U(1)$ transformation of the anomalous gauge boson $A_\mu \longrightarrow A_\mu + \partial_\mu \Lambda$, the variation of the effective action is non--zero.
The Green--Schwarz mechanism \cite{GS} provides a way to cancel these one--loop anomalies through the introduction of an antisymmetric $2$--form coupled at one--loop to the $U(1)_A$ $2$--form field strength, $F_{\rho \sigma }$, in the effective Lagrangian  
\begin{equation}
    - \frac{\zeta}{2}M_s ^2 \int d^4 x \; \epsilon^{\mu \nu \rho \sigma} B_{\mu \nu} F_{\rho \sigma },
\end{equation}
such that under a gauge transformation of the gauge field strength, the variation of the $B$ field compensates for the anomalous triangle diagram transformation. 

The computation of the Fayet--Iliopoulos coefficient $\zeta$ has been discussed in refs. 
\cite{FI1, FI2, FI3, FI4} 
and is performed by evaluating the $2$--point function of the antisymmetric $B$ field and the anomalous $U(1)_A$ gauge boson at one loop in the odd--spin structure, for chiral fermions charged under the anomalous $U(1)_A$ circulating in the loop. 
In order to soak up the zero modes of the ghost fields, one of the vertex operators has to be put in the $0$--picture, the other in the $-1$ and an insertion of the picture changing operator is needed, such that the zero modes are projected out of the integration.
The calculation can then be written as
\begin{equation}
\begin{split}
    \langle B^{\mu \nu}A^\rho \rangle & \sim \int_\mathcal{F}  \frac{d^2 \tau}{\tau_2} \int d^2z \: d^2w \: \langle V_{B,0} ^{\mu \nu} V_{A, -1} ^{\rho} e^{\phi} T_F  \rangle \\
    & \sim p_\alpha \int_\mathcal{F} \frac{d^2 \tau}{\tau_2} \int d^2z \: d^2w \: \langle \psi^\alpha \psi^\mu \psi^\rho \psi^ \sigma \rangle \langle \bar{\partial } X^\nu  \partial X^{\gamma} \rangle \langle \bar{J} \rangle \langle e^{-\phi} e^{\phi} \rangle   \eta_{\sigma \gamma}    .
\end{split}
\end{equation}
The computation is performed in the linear approximation $\mathcal{O}\left( p\right)$ evaluating the correlators at genus 1 each in its appropriate sector. Due to the fermion correlator, only the $\mathcal{N} = 1$ sector gives a non--vanishing result. The current correlator instead gives a contribution proportional to the $U(1)_A$ charges of the massless fermions in the loop. So the Fayet--Iliopoulos term reads
\begin{equation}
    \zeta = \frac{M_s ^2}{192\pi^2}\tr \left[ Q_A \right] , 
\end{equation}
where $M_s\approx g_s M_{\rm Planck}\approx g_s \cdot 5\times 10^{17}{\rm GeV}$ \cite{Kaplu}, 
and $Q_A$ are the charges of the matter states under the properly normalised 
$U(1)_A$.
In the numerical analysis in Section \ref{uplift} we will fix $g_s\sim {\rm O}(1)$. 
A more complete analysis would require some non--perturbative effect 
to stabilise the dilaton VEV that determines $g_s$. Such a mechanism may be induced by the
race--track mechanism \cite{Krasnikov}, or by a non--perturbative effect in $M$--theory \cite{Witgcu}.
Note in four dimensions the antisymmetric $B$ field can be dualized, on--shell, into the pseudoscalar axion field such that the coupling with the gauge and gravitational field strength terms, under a $U(1)_A$ gauge transformation, cancels the anomalous triangle diagram contribution.

When the trace $\tr [Q_A]$ is non--zero, an additional $D$--term appears in the potential of the form
\begin{equation}\label{VD}
    V_D = \frac{1}{2} g_s ^2 \zeta^2 .
\end{equation}

Even when the last supersymmetry is broken, this term still remains in the action and in particular gives an additional positive contribution to the minimum of the potential and eventually can uplift the minimum to a de Sitter one. 

In the heterotic string the $\mathcal{N}=4 \longrightarrow \mathcal{N} =1$ path is achieved by the introduction of the $\bm{b_1}$, $\bm{b_2}$ basis vectors associated to the $\mathbb{Z}_2 \times \mathbb{Z}_2$ orbifold (see Section \ref{PFSection}).
The way the last supersymmetry is broken depends on how we implement the breaking into the GGSO phases. These conditions are given in the following section in eq. (\ref{SUSYcond1}). An explicit breaking projects out the last gravitino of the spectrum, while in a spontaneous breaking induced by the Scherk--Schwarz mechanism \cite{SS1,SS2,SS3}, the gravitino acquires a mass and supersymmetry is restored at the boundary of the moduli space. 

One may question the notion of a $D$--term in non--supersymmetric string vacua, and in particular
its contribution to the vacuum energy. We may make an analogy with the gauge symmetries, which 
are broken directly at the string scale. The string spectra still preserves a memory of the 
underlying symmetries that play a role in {\it e.g.} the Yukawa coupling relations and 
flavour symmetries. Similarly, we may expect the string vacua to retain a memory of the 
underlying supersymmetric structures. In the case of the spontaneous Scherk--Schwarz 
breaking, supersymmetry is broken by coupling the boundary conditions of the superpartners 
of the internal dimensions to a shift in one of the compactified circles. In this case, 
the gravitino mass is proportional to the radius of the supersymmetry breaking circle and
we may indeed expect a $D$--term potential to be generated. In the cases with hard supersymmetry
breaking we may take the contribution to the vacuum energy on dimensional grounds. We note that 
the existence of a Fayet--Iliopoulos term in supergravity is an area of contemporary debate 
\cite{DienesThomas}, 
and therefore further scrutiny of our reasoning here is warranted. 

\section{Partition Function and One--Loop Potential}\label{PFSection}

We explore the one--loop cosmological constant and $U(1)_A$ tadpole calculations for models defined through the basis set
\begin{align}\label{SO10basis}
{\one}&=\{\psi^\mu,\
\chi^{1,\dots,6},y^{1,\dots,6}, \omega^{1,\dots,6}\ |   ~~~\overline{y}^{1,\dots,6},\overline{\omega}^{1,\dots,6},
\overline{\eta}^{1,2,3},
\overline{\psi}^{1,\dots,5},\overline{\phi}^{1,\dots,8}\},\nonumber\\
\Sv&=\{{\psi^\mu},\chi^{1,\dots,6}\},\nonumber\\
\e{i}&=\{y^{i},w^{i}\; | \; \overline{y}^i,\overline{w}^i\},
\
i=1,\dots,6,\nonumber\\
\bv{1}&=\{\chi^{34},\chi^{56},y^{34},y^{56}\; | \; \overline{y}^{34},
\overline{y}^{56},\overline{\psi}^{1,\dots,5},\overline{\eta}^1\},\\
\bv{2}&=\{\chi^{12},\chi^{56},y^{12},y^{56}\; | \; \overline{y}^{12},
\overline{y}^{56},\overline{\psi}^{1,\dots,5},\overline{\eta}^2\},\nonumber\\
\z{1}&=\{\overline{\phi}^{1,\dots,4}\},\nonumber\\
\z{2}&=\{\overline{\phi}^{5,\dots,8}\}.
\nonumber
\end{align}
Such a basis can be associated with symmetric $\mathbb{Z}_2\times \mathbb{Z}_2$ orbifolds \cite{z2z25} extensively classified in previous works (see  {\it e.g.} \cite{fknr,fkr1}) with an untwisted sector generating a gauge group of
\beq 
SO(10)\times U(1)_1\times U(1)_2\times U(1)_3\times SO(8)^2.
\eeq 

Models may then be defined through the choice of GGSO phases $C\smb{\bm{v_i}}{\bm{v_j}}$. There are 66 free phases for this basis, with all others specified by modular invariance. The full space of models is thus of size $2^{66}\sim 10^{19.9}$. Since we are interested in the non--supersymmetric vacua in this work we will be considering the set of vacua that project the potential gravitino arising from the $\bm{S}$ sector. The following GGSO phases can be fixed in order to retain $\mathcal{N}=1$ supersymmetry
\beq \label{SUSYcond1}
\CC{\mathds{1}}{\bm{S}}=\CC{\bm{S}}{\bm{S}}=\CC{\bm{S}}{\bm{e_i}}=\CC{\bm{S}}{\bm{b_k}}=\CC{\bm{S}}{\bm{z_1}}=\CC{\bm{S}}{\bm{z_2}}=-1
\eeq 
for $i=1,...,6$ and $k=1,2$. The space of non--supersymmetric vacua can then be explored by violating this condition.

\noindent
The generic form of the partition function for any model derived from the basis \eqref{SO10basis} can be written in a compact form as
\begin{aline}\label{PF}
    Z=\,&\frac{1}{\eta^{10}\bar{\eta}^{22}} \, \frac{1}{2^{2}} \sum_{\substack{a,k,\rho\\b,l,\sigma}} \;\frac{1}{2^{6}} \,\sum_{\substack{\zeta_i\\\delta_i}} \;\frac{1}{2^{4}}\sum_{\substack{h_1,h_2,H\\g_1,g_2,G}} (-1)^{a+b+ H G +\Phi\smb{a&k&\rho&\zeta_i&h_1&h_2&H}{b&l&\sigma&\delta_i&g_1&g_2&G}}\\[0.2cm]
    &\times \vth\smb{a}{b}_{\psi^\mu} \vth\smb{a+h_1}{b+g_1}_{\chi^{12}} \vth\smb{a+h_2}{b+g_2}_{\chi^{34}} \vth\smb{a-h_1-h_2}{b-g_1-g_2}_{\chi^{56}}\\[0.3cm]
    &\times \left| \vth\smb{\zeta_1}{\delta_1} \vth\smb{\zeta_1+h_1}{\delta_1+g_1} \vth\smb{\zeta_2}{\delta_2} \vth\smb{\zeta_2+h_1}{\delta_2+g_1} \right| \\[0.2cm]
    &\times \left| \vth\smb{\zeta_3}{\delta_3} \vth\smb{\zeta_3+h_2}{\delta_3+g_2} \vth\smb{\zeta_4}{\delta_4} \vth\smb{\zeta_4+h_2}{\delta_4+g_2} \right| \\[0.2cm]
    &\times \left| \vth\smb{\zeta_5}{\delta_5} \vth\smb{\zeta_5-h_1-h_2}{\delta_5-h_1-h_2} \vth\smb{\zeta_6}{\delta_6} \vth\smb{\zeta_6-h_1-h_2}{\delta_6-h_1-h_2} \right| \\[0.3cm]
     &\times \vthb\smb{k}{l}^5_{\bar{\psi}^{1-5}} \vthb\smb{k+h_1}{l+g_1}_{\bar{\eta}^1} \vthb\smb{k+h_2}{l+g_2}_{\bar{\eta}^2} \vthb\smb{k-h_1-h_2}{l-g_1-g_2}_{\bar{\eta}^3} \vthb\smb{\rho}{\sigma}^4_{\bar{\phi}^{1-4}} \vthb\smb{\rho + H}{\sigma + G}^4_{\bar{\phi}^{5-8}}.
\end{aline}
The modular invariant phase $\Phi\smb{a&k&\rho&\zeta_i&h_1&h_2&H}{b&l&\sigma&\delta_i&g_1&g_2&G}$ implements the various GGSO projections. A choice of phase is equivalent to a choice of GGSO matrix and hence there is a unique one--to--one map between them. The factor of $a+b$ ensures correct spin statistics, while the explicit inclusion of the extra phase $HG $ means that $\Phi=0$ is a valid modular invariant choice.

\noindent
The summation indices used to write the fermionic partition function \eqref{PF} correspond to various features of the model. The indices $a,b$ correspond to the spin structures of the spacetime fermions $\psi^\mu$, while $k,l$ are associated to the 16 right--moving complex fermions giving the gauge degrees of freedom of the heterotic string. The non--freely acting $\mathbb{Z}_2\times \mathbb{Z}_2$ orbifold twists are associated to the parameters $h_1,g_1$ and $h_2,g_2$. One of the key features of models defined by the basis \eqref{SO10basis} is the inclusion of the basis vectors $\e{i}$ which generate freely acting orbifold shifts in the internal dimensions of the compact torus. 
In order to render these shifts explicit in the partition function, we can introduce the twisted/shifted lattices $\Gamma^{(i)}_{2,2}$ of the underlying orbifold geometry that, 
at the maximally symmetric point $(T=i, U=\frac{i}{2})$, at which bosonic degrees of freedom can be fermionised, admit a factorised form which can be written entirely in terms of theta functions as follows
\begin{equation}\label{Lattice}
    \Gamma_{2,2}\smbar{H_1&H_2}{G_1&G_2}{h}{g} (i, \frac{i}{2}) = \frac{1}{4} \sum_{\zeta_i, \delta_i \in \mathbb{Z}}  \left| \vth\smb{\zeta_1}{\delta_1} \vth\smb{\zeta_1 + h}{\delta_1 + g} \vth\smb{\zeta_2}{\delta_2} \vth\smb{\zeta_2 + h}{\delta_2 +g} \right| (-1)^{(\zeta_i+h) G_i + (\delta_i+g) H_i +H_i G_i}
\end{equation}
Then, recasting appropriately the phase $\Phi$, the partition function (\ref{PF}) can be written as follows
\begin{aline}\label{PF2}
    Z=\,&\frac{1}{\eta^{10}\bar{\eta}^{22}} \, \frac{1}{2^{2}} \sum_{\substack{a,k,\rho\\b,l,\sigma}} \;\frac{1}{2^{6}} \,\sum_{\substack{H_i\\G_i}} \;\frac{1}{2^{4}}\sum_{\substack{h_1,h_2,H\\g_1,g_2,G}} (-1)^{a+b+ HG +\Phi\smb{a&k&\rho&H_i&h_1&h_2&H}{b&l&\sigma&G_i&g_1&g_2&G}}\\[0.2cm]
    &\times \vth\smb{a}{b}_{\psi^\mu} \vth\smb{a+h_1}{b+g_1}_{\chi^{12}} \vth\smb{a+h_2}{b+g_2}_{\chi^{34}} \vth\smb{a-h_1-h_2}{b-g_1-g_2}_{\chi^{56}}\\[0.3cm]
    &\times  \Gamma^{(1)}_{2,2}\smbar{H_1&H_2}{G_1&G_2}{h_1}{g_1} (i, \frac{i}{2}) \\[0.1cm]
    &\times  \Gamma^{(2)}_{2,2}\smbar{H_3&H_4}{G_3&G_4}{h_2}{g_2} (i, \frac{i}{2})  \\[0.1cm]
    &\times  \Gamma^{(3)}_{2,2}\smbar{H_5&H_6}{G_5&G_6}{h_1+h_2}{g_1+g_2} (i, \frac{i}{2})  \\[0.3cm]
    &\times \vthb\smb{k}{l}^5_{\bar{\psi}^{1-5}} \vthb\smb{k+h_1}{l+g_1}_{\bar{\eta}^1} \vthb\smb{k+h_2}{l+g_2}_{\bar{\eta}^2} \vthb\smb{k-h_1-h_2}{l-g_1-g_2}_{\bar{\eta}^3} \vthb\smb{\rho}{\sigma}^4_{\bar{\phi}^{1-4}} \vthb\smb{\rho + H}{\sigma + G}^4_{\bar{\phi}^{5-8}}.
\end{aline}
Now the indices $H_i,G_i$ parameterize each of the six independent shifts. The additional indices $\rho, \sigma$ and $H,G$ correspond to the basis vectors $\z{1}$ and $\z{2}$ acting on the hidden sector of our model. \\

\noindent
The form of the twisted/shifted lattice dependent on the moduli $T^{(i)}$ and $U^{(i)}$ of the compact $\mathbf{T}^6=\mathbf{T}^2\times \mathbf{T}^2 \times \mathbf{T}^2$ requires closer attention. We know that all dependence on the geometric moduli is contained in the untwisted sector of the model and hence
\begin{equation}\label{Z22Twisted}
    \Gamma_{2,2}\smbar{H_1&H_2}{G_1&G_2}{h}{g}(T,U)\Big|_{h,g\neq0} = \Gamma_{2,2}\smbar{H_1&H_2}{G_1&G_2}{h}{g}(T_*,U_*).
\end{equation}
This means that for nonzero twists the lattice is precisely given by its factorised form in (\ref{Lattice}). Here $T=T_1+iT_2$, $U=U_1+iU_2$ are the moduli of the torus, and parameterise the metric and the antisymmetric tensor field of the two dimensional torus and $\left(T_*, U_*=i, i/2 \right)$. In the case of the untwisted sector, the shifted lattice can be written in a Poisson resummed Hamiltonian form as
\begin{equation}\label{Z22TU}
    \Gamma_{2,2}\smbar{H_1&H_2}{G_1&G_2}{0}{0} (T,U) = \sum_{m_i, n_i \in\mathbb{Z}} q^{\frac{1}{2} |\mathcal{P}_L(T,U)|^2} \bar{q}^{\frac{1}{2} |\mathcal{P}_R(T,U)|^2}  e^{i \pi (G_1 m_1 + G_2 n_2)}
\end{equation}
where the left and right--moving momenta are 
\begin{aline}\label{PLPR}
    \mathcal{P}_L =& \frac{1}{\sqrt{2 T_2 U_2}} \left[
    m_2 + \frac{H_2}{2} - U m_1 +T \left( n_1 + \frac{H_1}{2} + U n_2 \right) \right] \\
    \mathcal{P}_R =& \frac{1}{\sqrt{2 T_2 U_2}} \left[
    m_2 + \frac{H_2}{2} - U m_1 +\bar{T} \left( n_1 + \frac{H_1}{2} + U n_2 \right) \right].
\end{aline}
Written in this form, it is easy to extract the $q$--expansion of the partition function at any given point in the moduli space which is crucial for calculating the one--loop potential. It can be shown that the twisted/shifted lattice sums \eqref{Z22Twisted} and \eqref{Z22TU} evaluated at the special point $(T_*,U_*)$ indeed reproduce the free fermionic form of the partition function \eqref{PF2}. 
\\

Given the partition function \eqref{PF2}, the one--loop potential is evaluated by summing over all inequivalent worldsheet tori via the modular invariant integral
\begin{equation}\label{PotentialTU}
    V_\text{one--loop}(T^{(i)},U^{(i)}) = -\frac{1}{2}\frac{M_s^4}{(2\pi)^4} \int_\mathcal{F}\frac{d^2\tau}{\tau_2^2}\, Z(\tau,\bar{\tau},T^{(i)},U^{(i)}),
\end{equation}
where in $ Z(\tau,\bar{\tau},T^{(i)},U^{(i)})$ we have now taken into account the extra two bosonic degrees of freedom arising from the worldsheet. In models with an anomalous $U(1)$, an additional contribution to the potential $V_D$ is generated as discussed in Section \ref{AnomU1}. Since this term is independent of the geometric moduli it provides a constant shift of the potential throughout moduli space. Hence we can write a total potential as
\begin{equation}\label{VUplift}
    V_\text{total}=V_\text{one--loop}(T^{(i)},U^{(i)})+V_D,
\end{equation}
where $V_D$ is given in term of the trace of the anomalous $U(1)$ via \eqref{VD}.

\section{The Uplifted String Model}\label{uplift}

The anomalous $U(1)_A$ in the string models generated by the basis vectors in 
eq. (\ref{SO10basis}), when non--vanishing, is given by 
\beq 
U(1)_A=
\sum_{i=1}^3 \frac{aU_1+bU_2+cU_3}{\sqrt{a^2+b^2+c^2}}
\eeq 
where $(a,b,c)=\frac{1}{k}(\tr[U_1],\tr[U_2],\tr[U_3])$ and $k=\text{gcd}(\tr[U_1],\tr[U_2],\tr[U_3])$,
with $U_i=U(1)_i$ being generated by the world--sheet currents $: \bar{\eta}^{i*} \bar{\eta}^i :$, for $i=1,2,3$. 
We note that with the set of basis vectors given in eq. (\ref{SO10basis}), $U(1)_{1,2,3}$ are the 
only $U(1)$ symmetries that are left unbroken in the four--dimensional gauge group. This may, in general,
be different in models that utilise asymmetric boundary conditions, and in which the 
$SO(10)$ symmetry is broken to a subgroup. In the first case, there are additional boundary conditions
arising from the internal compactified space, whereas in the second there may be additional $U(1)$
symmetries arising from the hidden sector \cite{fny,slm2}. 

To demonstrate the possibility of using the Fayet--Iliopoulos D--term to uplift the one--loop potential we take the following GGSO configuration
{\begin{equation}
\small
\CC{\bm{v_i}}{\bm{v_j}}= 
\begin{blockarray}{cccccccccccccc}
&\mathbf{1}& \Sv & \e{1}&\e{2}&\e{3}&\e{4}&\e{5}&\e{6}&\bv{1} & \bv{2} &\z{1} & \z{2} \\
\begin{block}{c(rrrrrrrrrrrrr)}
\mathbf{1}&-1&-1& 1& 1&-1& 1&-1& 1& 1& 1&-1& 1&\ \\ 
\Sv       &-1&-1&-1&-1&-1&-1&-1&-1& 1&-1&-1& 1&\ \\
\e{1}     & 1&-1&-1&-1&-1& 1& 1& 1& 1& 1&-1&-1&\ \\ 
\e{2}     & 1&-1&-1&-1&-1&-1&-1&-1& 1&-1& 1& 1&\ \\ 
\e{3}     &-1&-1&-1&-1& 1&-1&-1&-1&-1&-1&-1&-1&\ \\
\e{4}     & 1&-1& 1&-1&-1&-1& 1& 1&-1&-1&-1&-1&\ \\
\e{5}     &-1&-1& 1&-1&-1& 1& 1&-1& 1&-1&-1& 1&\ \\ 
\e{6}     & 1&-1& 1&-1&-1& 1&-1&-1& 1&-1&-1&-1&\ \\ 
\bv{1}    & 1&-1& 1& 1&-1&-1& 1& 1& 1& 1& 1&-1&\ \\ 
\bv{2}    & 1& 1& 1&-1&-1&-1&-1&-1& 1& 1& 1&-1&\ \\ 
\z{1}     &-1&-1&-1& 1&-1&-1&-1&-1& 1& 1&-1&-1&\ \\
\z{2}     & 1& 1&-1& 1&-1&-1& 1&-1&-1&-1&-1& 1&\ \\
\end{block}
\end{blockarray} \ .
\label{ggsophases}
\end{equation}}

\noindent The modular invariant phase corresponding to this choice is given by

\begin{aline}
    \Phi\smb{a&k&\rho&H_i&h_1&h_2&H}{b&l&\sigma&G_i&g_1&g_2&G} =&
    ~~b \left( a + H + h_2 + k + \rho \right)\\
    & + l \left( a + H + h_2 + H_2 + H_3 + H_5 + H_6 + \rho \right)\\
    & + \sigma \left( a + H + h_1 + h_2 + H_2 + H_3 + H_6 + k + \rho \right)\\
    & + G_1 \left( h_2 + H_2 + H_3 \right)\\
    & + G_2 \left( H_1 + H_4 + H_5 + k + \rho \right)\\
    & + G_3 \left( H + H_1 + h_2 + H_4 + H_5 + k + \rho \right)\\
    & + G_4 \left( h_1 + H_2 + H_3 \right)\\
    & + G_5 \left( H_2 + H_3 + H_5 + H_6 + k \right)\\
    & + G_6 \left( H + h_1 + h_2 + H_5 + k + \rho \right)\\
    & + g_1 \left( h_1 + h_2 + H_4 + H_6 + \rho \right)\\
    & + g_2 \left( a + H + h_1 + H_1 + h_2 + H_3 + H_6 + k + \rho \right)\\
    & + G \left( a + H + h_2 + H_3 + H_6 + k + \rho \right),
\end{aline}
\noindent where we note that the breaking of supersymmetry is induced by the term $al+a\sigma+aG + bk +b\rho +bH$ which corresponds to the GGSO phase $C\smb{\mathbf{S}}{\mathbf{z_2}}=+1$.\\
\noindent
According to the discussion in Section \ref{PFSection}, we can choose to analyse the behaviour of the vacuum energy in all or some directions of the geometric moduli space parametrised by the $T^{(i)}, U^{(i)}$. Evaluating the behaviour of the potential over the entire moduli space is beyond the scope of this paper, however, the obstructions to doing so are purely based on computational time constraints and the techniques described above are general. A usual choice to make is the volume of the first torus $T^{(1)}_2$ as done so in \cite{fr1,fr2}. It is important to note, however, that this choice is somewhat arbitrary. $T_2$ is usually chosen as in the case of a Scherk--Schwarz breaking of supersymmetry, one can choose a configuration in which the scale of the breaking is controlled by the volume of the first torus. As we will demonstrate in what follows, our model corresponds to a hard breaking of supersymmetry and hence such a justification is not valid. 

\begin{figure}[t]
\centering
\includegraphics[width=0.9\linewidth]{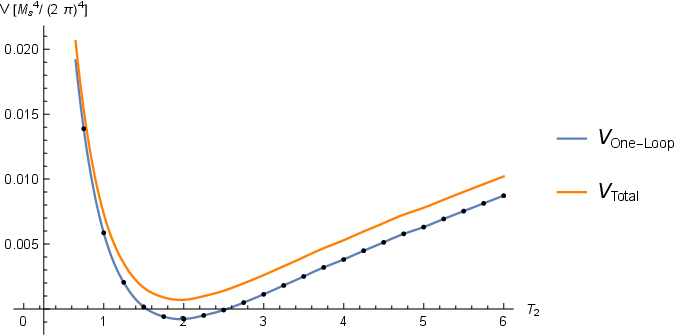}
\caption{ One--loop Potential of Example Model before and after uplift by the FI contribution via \eqref{VUplift}. }
\label{UpliftedPot}
\end{figure}

Starting with the model defined by the basis vectors \eqref{SO10basis} and GGSO configuration \eqref{ggsophases}, the theory can be deformed in the $T_2$ direction by implementing the moduli dependence via the replacement
\begin{equation}
    \Gamma^{(1)}_{2,2}\smbar{H_1&H_2}{G_1&G_2}{h_1}{g_1}(T^{(1)}_*,U^{(1)}_*)  \longrightarrow \Gamma^{(1)}_{2,2}\smbar{H_1&H_2}{G_1&G_2}{h_1}{g_1}(T_2^{(1)},T_{1*}^{(1)},U^{(1)}_*)
\end{equation}
in the partition function \eqref{PF2}. This means that we fix all other geometric moduli at the free fermionic point while varying $T^{(1)}_2$ freely. This produces a model with a minimum for the potential at $T_2=2$ where the one--loop cosmological constant takes the value $\Lambda=-0.000785598 \, \mathcal{M}^4$ as shown in Figure \ref{UpliftedPot}. As per Section \ref{AnomU1}, the trace of the anomalous $U(1)$ for this model is $\tr[U(1)_A]=144/\sqrt{2}$ which generates a FI contribution of $V_D=0.00144365 \, \mathcal{M}^4$ to the potential ensuring a positive uplifted minimum via \eqref{VUplift} as depicted in Figure \ref{UpliftedPot}.

\section{Conclusion}

String theory provides a self--consistent framework for the synthesis of gravity and quantum 
mechanics. String phenomenology aims to connect string theory with observational data. For 
that purpose, detailed phenomenological models were constructed.
The free fermionic models, which correspond to $\mathbb{Z}_2\times \mathbb{Z}_2$ toroidal orbifold compactifications
at special points in the moduli space, 
provide a large space of three--generation models with an unbroken $SO(10)$ subgroup that 
can be further broken to the Standard Model in the effective field theory limit.
The majority of these constructions possess $\mathcal{N}=1$ spacetime
supersymmetry in four dimensions. 
Making contact with observational data mandates moving away from the stability afforded by 
supersymmetry. For that purpose, over the past few years a systematic classification program 
of tachyon--free non--supersymmetric string models was developed. 

A recurring feature in the $\mathcal{N}=1$ string models is the existence of an anomalous $U(1)$ gauge symmetry
that generates a non--trivial Fayet--Illiopoulos $D$--term that breaks supersymmetry. Supersymmetry
is restored by giving Vacuum Expectation Values (VEVs) to some Standard Model singlets in the 
massless string spectrum, along $F$-- and $D$--flat directions, that restores $\mathcal{N}=1$ supersymmetry
at the string scale. Supersymmetry is then expected to be broken by some non--perturbative effect, 
{\it e.g.} by hidden sector gaugino condensation. Restoration of $\mathcal{N}=1$ supersymmetry 
implies that the vacuum energy at the string scale vanishes in these models. 

Anomalous $U(1)$ symmetries also arise in non--supersymmetric string vacua, and the same 
diagrams that lead to the Fayet--Illiopoulos $D$--term in $\mathcal{N}=1$ supersymmetric models 
are generated in the non--supersymmetric models. Hence, similar contributions to the vacuum energy arise
in these non--supersymmetric configurations and their effects have to be taken into account.
The possibility then exists that the would--be $D$--term contribution lifts an \textit{a priori} negative 
vacuum energy to a positive value. We discussed this scenario in Section \ref{uplift}, where the one--loop vacuum energy, as well as 
the would--be $D$--term contribution,
is calculated in a specific heterotic string model. 
This possibility was envisioned by Burgess, Kallosh and Quevedo \cite{BKQ}. 
We emphasise, however, that the analysis in Section \ref{uplift} is for illustration 
purposes only. Indeed, there are many issues that have not been addressed, including the stabilisation
of the dilaton and the other moduli in the string vacuum as well as the backreaction on the 
internal and spacetime geometry. Such issues have to be addressed before an informed statement 
can be made about the existence of stable string vacua with positive cosmological constant. 

Nevertheless, the would--be $D$--term is prevalent in non--supersymmetric string vacua and 
its contribution has to be taken into account. Non--supersymmetric string vacua include 
those that correspond to compactifications of the $SO(16)\times SO(16)$ heterotic string, 
as well as those that correspond to the tachyon--free compactifications of the tachyonic ten--dimensional configurations. In the first class, we can distinguish between string vacua in which
supersymmetry is broken by a Scherk--Schwarz mechanism versus those in which it is broken 
explicitly. In the first class, we expect supersymmetry to be restored when the radius of the 
Scherk--Schwarz circle goes to infinity, whereas in the second it does not. A more detailed 
analysis of the different cases will be presented in a forthcoming publication \cite{DFMP2}. 
We note that the model in Section \ref{uplift} is of the second type.
We further remark that many of the geometrical moduli in the free fermionic string 
models can be fixed by using asymmetric boundary conditions and such configurations offer
a more restricted framework to investigate the issue of stability. 

\medskip
\section*{Acknowledgments}
AEF would like to thank the CERN theory division for hospitality and support.
The work of ARDA is supported in part by EPSRC grant EP/T517975/1.

\printbibliography[heading=bibintoc]

\end{document}